\documentclass[prl,twocolumn,amsmath,nobibnotes,nofootinbib,superscriptaddress,preprintnumbers,hyperref]{revtex4-1}

\usepackage{bm} \usepackage{graphicx} \usepackage{color}
\usepackage{epstopdf} \usepackage{amssymb}
\def\ba{\begin{eqnarray}}
\def\ea{\end{eqnarray}}
\def\be{\begin{equation}}
\def\ee{\end{equation}}
\def\({\left(}
\def\){\right)}
\def\[{\left[}
\def\]{\right]}
\def\<{\left<}
\def\>{\right>}

\newcommand{\prob}
  {{\rm{Pr}}}
\newcommand{\ns}
  {{N_{\rm{s}}}}
\newcommand{\nblob}
  {{N_{\rm{b}}}}
\newcommand{\nsavge}
  {{\bar{N}_{\rm{s}}}}
  
  \newcommand{\src}
  {s}

\newcommand{\template}
  {\mathbf{t}}
\newcommand{\model}
  {\mathbf{m}}
  \newcommand{\fsky}
  {f_{\rm{sky}}}
  \newcommand{\data}
  {\mathbf{d}}
  
	\newcommand{\diff}
  {{\rm{d}}}
  \newcommand{\matr}[1]
        {\mbox{\bf \sf{#1}}}
  
  \newcommand{\blob}
  {b}

\begin{document}

\title{First Observational Tests of Eternal Inflation}
\date{\today}

\author{Stephen M. Feeney}
\email{stephen.feeney.09@ucl.ac.uk}
\affiliation{Department of Physics and Astronomy, University College London, London WC1E 6BT, U.K.}
\author{Matthew C. Johnson}
\email{mjohnson@perimeterinstitute.ca}
\affiliation{Perimeter Institute for Theoretical Physics, Waterloo, Ontario N2L 2Y5, Canada} 
\affiliation{California Institute of Technology, Pasadena, CA 91125, USA}
\author{Daniel J. Mortlock}
\email{mortlock@ic.ac.uk}
\affiliation{Astrophysics Group, Imperial College London, Blackett Laboratory, Prince Consort Road, London, SW7 2AZ, U.K.}
\author{Hiranya V. Peiris}
\email{h.peiris@ucl.ac.uk}
\affiliation{Department of Physics and Astronomy, University College London, London WC1E 6BT, U.K.}
\affiliation{Institute of Astronomy and Kavli Institute for Cosmology, University of Cambridge, Cambridge CB3 0HA, U.K.}

\begin{abstract}
The eternal inflation scenario predicts that our observable universe resides inside a single bubble embedded in a vast inflating multiverse. We present the first observational tests of eternal inflation, performing a search for cosmological signatures of collisions with other bubble universes in cosmic microwave background data from the WMAP satellite. We conclude that the WMAP 7-year data do not warrant augmenting $\Lambda$CDM with bubble collisions, constraining the average number of detectable bubble collisions on the full sky $\nsavge < 1.6$ at $68 \%$ CL. Data from the {\em Planck} satellite can be used to more definitively test the bubble collision hypothesis.
\end{abstract}

\preprint{}

\maketitle

{\bf Introduction:} The inflationary paradigm has been very successful at explaining the initial conditions giving rise to our observable universe. Considering the initial conditions for inflation itself leads to the possibility that our observable universe might only be a tiny piece of a vast multiverse. In this scenario, known as eternal inflation (for a review, see e.g. Ref.~\cite{Aguirre:2007gy}), our observable universe resides inside a single bubble nucleated out of a false vacuum de Sitter space. The rate of bubble formation is outpaced by the accelerated expansion of the inflating false vacuum, and therefore inflation does not end everywhere. 

Eternal inflation is ubiquitous in theories with extra dimensions (string theory being the primary example) and positive vacuum energy. However, testing this scenario is extremely difficult since eternal inflation is a {\em pre-inflationary} epoch: any signals from outside of our bubble would naively appear to be stretched to unobservable super-horizon scales. While this is in general true, one prospect for probing this epoch lies in the observation of the collisions between vacuum bubbles. These collisions produce inhomogeneities in the inner-bubble cosmology, raising the possibility that their effects are imprinted in the cosmic microwave background (CMB)~\cite{Aguirre:2007an}. In this paper we describe a robust algorithm designed to test the hypothesis that there are bubble collisions in the {\em Wilkinson Microwave Anisotropy Probe} (WMAP) 7-year data~\cite{Jarosik:2010iu}. More generally, our analysis pipeline can be adapted to test any theory predicting localized signatures on the CMB sky. 

Recent theoretical work (see the review Ref.~\cite{Aguirre:2009ug} and references therein) has established that bubble collisions can produce detectable signals and still be compatible with our observed cosmology; in addition, there are models that predict an expected number of {\em detectable} collisions $\nsavge$ larger than one. The value of $\nsavge$  is highly dependent on: the scalar field potential(s) that drive eternal inflation (which controls the rate of bubble formation); the duration of inflation inside our bubble (the more inflation the weaker the signal); and the particular realization of the CMB sky and bubble collisions we might observe (i.e. even a strong signal could be obscured by foregrounds). Although there is ample motivation to consider the eternal inflation scenario, a concrete model (of, say, the string theory landscape) providing all of these details does not currently exist.

Nevertheless, a single bubble collision produces a rather generic set of signatures that we target in our analysis:\\
{\bf{\em Azimuthal symmetry:}} A collision between two bubbles leaves an imprint on the CMB sky that has azimuthal symmetry. This is a consequence of the SO(2,1) symmetry of the spacetime describing the collision of two vacuum bubbles.\\
{\bf{\em Causal boundary:}} The surface of last scattering can only be affected inside the future light cone of a collision event, which forms a ring on the CMB sky. The observed temperature of the CMB need not be continuous across this boundary. \\
{\bf{\em Long-wavelength modulation inside the causal boundary:}} A bubble collision is a pre-inflationary relic. The effects of a collision are stretched by inflation, and induce an overall modulation of the CMB temperature anisotropies~\cite{Chang:2008gj}.

\begin{figure}
   \includegraphics[width=8.5 cm]{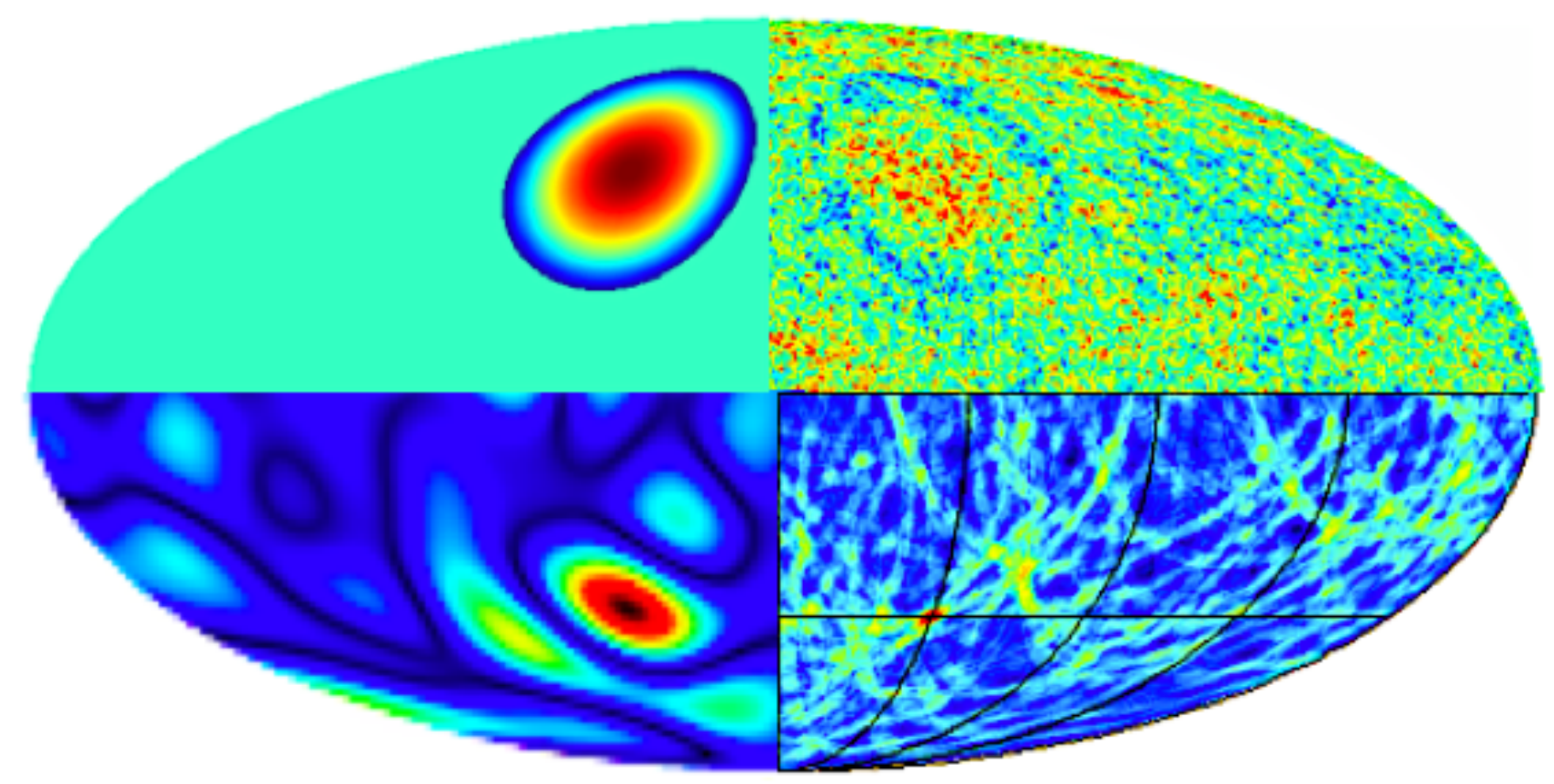}
\caption{The signatures of a bubble collision at various stages in our analysis pipeline. A collision (top left) induces a temperature modulation in the CMB temperature map (top right). The ``blob" associated with the collision is identified by a large needlet response (bottom left), and the presence of an edge is highlighted by a large response from the edge detection algorithm (bottom right). In parallel with the edge-detection step, we perform a Bayesian parameter estimation and model selection analysis.}
\label{fig-multiplebubble}
\end{figure}

At the time of last scattering, the signal is long-wavelength, has approximate planar symmetry (from the small observed value of the normalized curvature density, $\Omega_k$), and has a causal boundary. The observed temperature fluctuation (to lowest order in $\cos \theta$, centered on the north pole) takes the form
\begin{equation}\label{eq:tempmod}
\frac{\delta T}{T} = \left[  \frac{z_{\rm crit} - z_0 \cos \theta_{\rm crit}}{1 - \cos \theta_{\rm crit}} + \frac{z_0 - z_{\rm crit}}{1 - \cos \theta_{\rm crit}} \cos \theta \right] \Theta (\theta_{\rm crit} - \theta),
\end{equation}
where $z_0$ is the amplitude of the temperature modulation at the centre of the collision, $z_{\rm crit}$ is the discontinuity of the temperature at the causal boundary $\theta = \theta_{\rm crit}$, and $\Theta (\theta_{\rm crit} - \theta)$ is a step function. An example is shown in the top left of Fig.~\ref{fig-multiplebubble}. A temperature modulation of this form, with $z_{\rm crit} = 0$, was first derived by Chang et. al.~\cite{Chang:2008gj}, who  analyzed the behaviour of the inflaton field in a background thin-wall collision geometry. Our model allows, but does not require, the existence of a temperature discontinuity at the edge. Including the location of the collision centre, $\{ \theta_0, \phi_0 \}$, a collision can be described by five parameters: $\{ z_0, z_{\rm crit}, \theta_{\rm crit}, \theta_0, \phi_0 \}$. A complete model of eternal inflation would predict a probability distribution for the number of collisions and their parameters. In the absence of such specific predictions we use $\nsavge$ as a continuous parameter that characterizes particular models of eternal inflation. The concordance $\Lambda$CDM cosmological model is given by the special case in which $\nsavge = 0$. 

{\bf Data and simulations:} We calibrate our analysis pipeline and determine its sensitivity using simulations of the WMAP 7-year data. WMAP has measured the intensity and polarization of the microwave sky in five frequency bands. We perform our final analysis on the foreground-subtracted 94 GHz W-band WMAP temperature map, as this combines high resolution (determined by the detector's $0.22^\circ$ beam) with small foreground contamination. To minimize the effects of the residual foregrounds we cut the sky with the conservative KQ75 mask, leaving 70.6\% of the sky unmasked. 

The WMAP data-reduction and foreground-removal pipelines may leave behind systematic errors that we are unable to include in our likelihood function. To calibrate our detection thresholds and quantify the expected false detection rate due to such systematic effects, we use an end-to-end simulation of the W-band channel of the WMAP instrument, provided by the WMAP Science Team. This is generated from a simulated time ordered data stream (containing Galactic foregrounds, CMB fluctuations, realistic instrumental noise, finite beam size, and other instrumental effects) which is processed using the same pipeline applied to the actual data. 

To determine the sensitivity of our analysis pipeline to bubble collisions, we generate simulated temperature maps containing CMB fluctuations, a bubble collision, instrumental noise, and smoothing to emulate the W-band beam. We consider collisions with $\theta_{\rm crit} = 5^{\circ}, 10^{\circ}, 25^{\circ}$ and choose centres in a high-  and low-noise unmasked region of the sky. For each $\theta_{\rm crit}$ and location, we generate $35$ simulated collisions with parameters logarithmically distributed in the ranges $10^{-6} \leq z_{0} \leq 10^{-4}$ and $-10^{-4} \leq z_{\rm crit} \leq -10^{-6}$; the lower limit of the $z_{\rm crit}$ parameter range produces templates with edges which are undetectable in WMAP data. The response of our pipeline depends only on the absolute value of $z_0$ and $z_{\rm crit}$, so the choice of sign for $z_0$ and $z_{\rm crit}$ is arbitrary. We repeat this process for three realizations of the background CMB fluctuations, yielding a total of $210$ simulated sky maps for each of the three collision sizes.

{\bf Method:} Our primary goal is to determine, given the WMAP 7-year data, what constraints can be placed on $\nsavge$ and whether models predicting $\nsavge > 0$ should be be preferred over models predicting $\nsavge = 0$. From Bayes' theorem (e.g. Ref.~\cite{Sivia_Skilling:2006}), this information is contained in the posterior probability distribution for $\nsavge$ given full-sky CMB data. Evaluating this would require the computationally intractable tasks of inverting the full-sky full-resolution CMB covariance matrix and integrating the bubble-collision likelihood over a many-dimensional parameter space. However, as we describe in more detail in a companion paper~\cite{Feeney:2010dd}, it is possible to approximate the full-sky Bayesian evidence integral by a patch-wise analysis, if it is possible to first identify the regions on the sky that contribute the most to the integrand. Our analysis pipeline automatically locates these regions, avoiding {\em a posteriori} selection effects~\cite{Bennett:2010jb}. Once a set of candidate patches has been identified, it is also possible to apply further tests of the data in parallel.

Our analysis pipeline is schematically depicted in Fig.~\ref{fig-multiplebubble}. A ``blob detection" step using needlets~\cite{Marinucci:2007aj} (a type of wavelet) is used to identify significant features in the temperature map and determine their approximate location and angular size. Two parallel verification steps are then performed: first, an edge detection algorithm is applied to these features to search for circular temperature discontinuities ($z_{\rm crit}$ in the template). Second, a pixel-based Bayesian model selection and parameter estimation analysis is performed on the regions of the sky highlighted by the needlets. These results are used to construct an approximation to the full-sky posterior distribution of $\nsavge$, from which we can determine if the data supports augmenting $\Lambda$CDM with bubble collisions.

{\bf{\em Blob detection:}} To identify the most promising candidate signals in a temperature map, we perform a suite of spherical needlet transforms
\begin{equation}\label{eq:needlettransform}
\beta_{jk} = \sum_{\ell}b_{\ell}(j)\sum^{\ell}_{m=-\ell}a_{\ell m}Y_{\ell m}(\xi_{jk}),
\end{equation}
where $j =$ frequency, $k =$ HEALPix pixel \cite{Gorski:2004by}, $Y_{\ell m}$ are the spherical harmonics, $a_{\ell m}$ are the spherical harmonic coefficients of an input map, and $b_{\ell}(j)$ is a frequency-dependent filter function determining the needlet shape. Two classes of  needlet shapes (``standard'' \cite{Marinucci:2007aj} and ``Mexican'' \cite{Scodeller:2010mp}) are used to ensure that we are sensitive to a wide variety of modulations. We optimize the needlet response to a variety of collision templates by adjusting the band-width of $b_{\ell}(j)$. Optimal unbiased maximum-likelihood estimators of the $a_{\ell m}$s~\cite{deOliveiraCosta:2006zj} are used at low $\ell$ to partially offset the effects of the mask on low-frequency needlet coefficients;  cut-sky $a_{\ell m}$s are used at larger $\ell$.  

In the lower left quadrant of Fig.~\ref{fig-multiplebubble} we plot $\beta_{jk}$ for fixed $j$ at each pixel $k$ obtained for a simulated map containing a collision. The needlet coefficients clearly take their largest value in the vicinity of the collision. Finding the frequency of maximum needlet response yields information about the angular scale of the collision.

For a purely Gaussian uncut CMB sky, the average needlet coefficient is zero for all $j,k$, and the variance of the needlet coefficients for fixed $j$ at each pixel $k$ is identical and directly related to the scalar temperature power spectrum. Cutting the sky introduces a $j$- and $k$-dependent bias, and we determine the significance of a needlet coefficient by 
\begin{equation}\label{eq:needletsignificance}
S_{jk} = \frac{|\beta_{jk} - \langle\beta_{jk}\rangle_{\mathrm{gauss, cut}}|}{\sqrt{\langle\beta^{2}_{jk}\rangle_{\mathrm{gauss, cut}}}},
\end{equation}
where the average $\langle\beta_{jk}\rangle_{\mathrm{gauss, cut}}$ and variance $\langle\beta^{2}_{jk}\rangle_{\mathrm{gauss, cut}}$ are calculated at each pixel from the needlet coefficients of 3000 collision-free Gaussian CMB realizations with the WMAP 7-year KQ75 sky cut applied.

Because there are many independent needlet coefficients in any given realization, there can be features of fairly large significance even in a purely Gaussian map. In addition, residual foregrounds and artifacts of experimental systematics can be picked up by the needlet transform. To minimize the number of such false detections, we use the needlet response to the end-to-end simulation to define a set of frequency- and shape-dependent detection thresholds (in the range $3.00 \le S_{jk} \le 3.75$). Regions with five or more pixels whose needlet coefficients exceed these thresholds, and which do not intersect the Galactic mask, are grouped into ``blobs", and are passed to the following steps. A total of 10 false detections are found in the end-to-end simulation using these thresholds. The angular scale of these features is read from a lookup table, built by finding the range of $ \theta_{\rm crit}$ over which each needlet frequency yields the maximum response to a set of simulated collision templates

\begin{figure*}[tb]
   \includegraphics[width=6 cm]{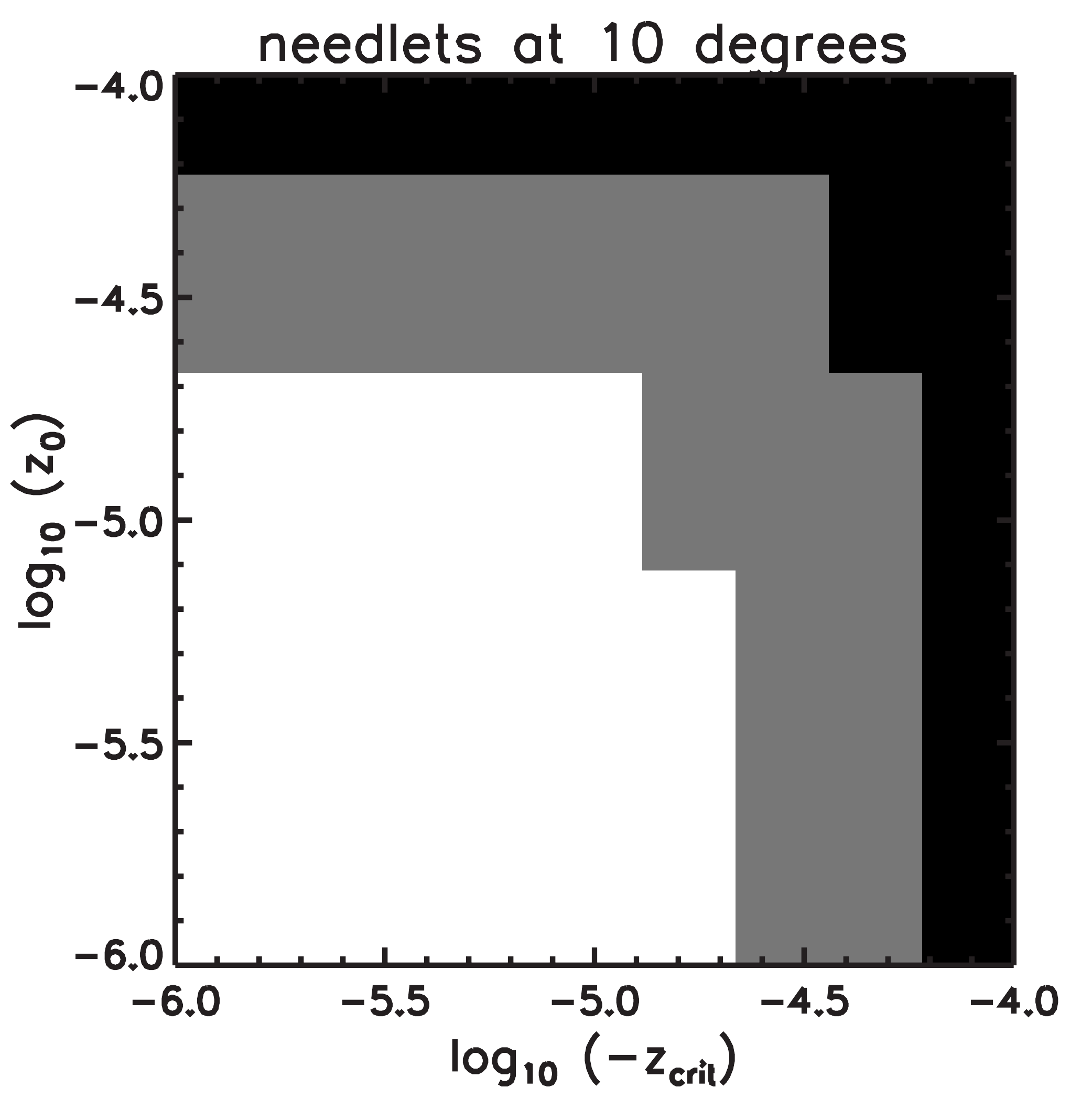}
   \includegraphics[width=6 cm]{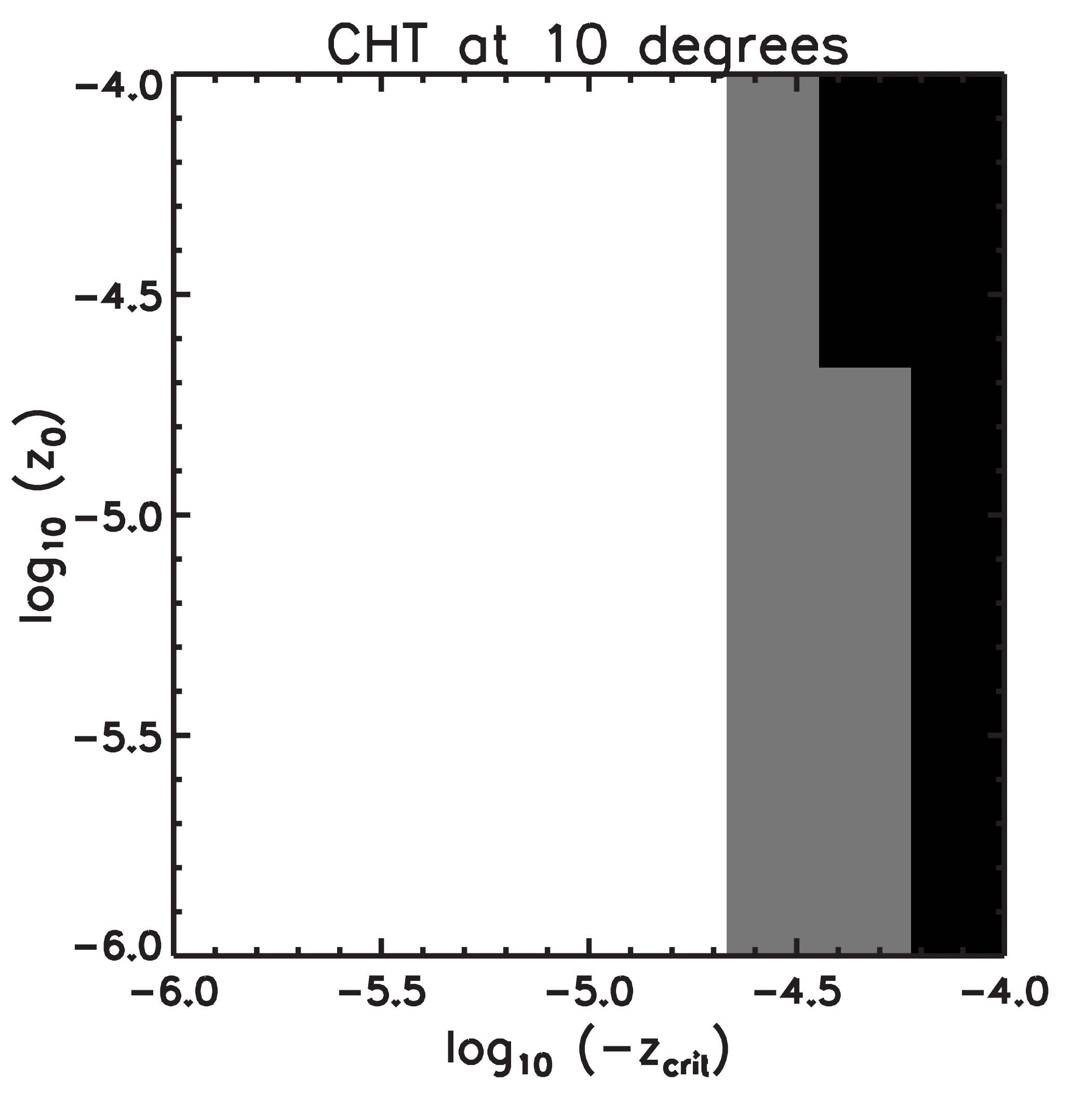}
\caption{Exclusion (black) and sensitivity (grey) regions for the needlet (left) and edge-detection (right) steps of the analysis pipeline applied to a set of $\theta_{\rm crit} = 10^{\circ}$ simulated bubble collisions. Collisions in the exclusion region would be definitively detected as long as they were not significantly masked. Collisions in the sensitivity region could be found if they were in a favorable location of the sky (i.e. low noise, or a region with a specific realization of CMB fluctuations which did not obscure the causal edge).}
\label{fig-exclusionplots}
\end{figure*}

We have tested the performance of the blob detection step using the full set of simulated collisions. The results are presented in Fig.~\ref{fig-exclusionplots} (left panel) as ``exclusion" and ``sensitivity" regions of the $\{ z_0, z_{\rm crit}\}$ parameter space for $\theta_{\rm crit} = 10^{\circ}$ (the $\theta_{\rm crit} = 5^{\circ}$ and $25^{\circ}$ results are nearly identical).  If we detect all six simulated bubble collisions at a given point $\{z_0, z_{\rm crit}\}$ of parameter space, then a non-detection in the real data would imply we can exclude such collisions; if we detect a collision for a subset of the simulations, then we are sensitive to collisions in this region of parameter space. 

{\bf{\em Edge detection:}} This step of the pipeline tests features highlighted by the blob detection stage for circular temperature discontinuities (controlled by $z_{\rm crit}$). To generate a set of candidate edges, we have developed an implementation of the Canny algorithm \cite{11275} for the HEALPix pixelization scheme. In this algorithm, the gradient of an image is generated, smoothed to reduce the effects of pixel noise, and thinned into local maxima, the strongest of which are stitched together into edges. 

An adaptation of the Circular Hough Transform (CHT) algorithm~\cite{360677} is then used to assess whether these candidate edges fall on circles of varying centre and angular radius. If a map of the ``CHT score" (the fraction of pixels in an annulus surrounding a given center that are candidate edges) is sharply peaked at some angular scale $\theta$, then there is evidence for a circular edge of this radius centered on the peak. In the example in the bottom right quadrant of Fig.~\ref{fig-multiplebubble}, the peak in the CHT score at the true centre of the collision shows an unambiguously detected edge. We have verified that none of the false detections flagged in the end-to-end simulations have a significant peak.

The sensitivity and exclusion regions from the edge-detection step are shown in Fig.~\ref{fig-exclusionplots} (right panel) for the $10^{\circ}$ simulated collisions. For the $5^{\circ}$ collisions, the proliferation of degree-sized features in the background CMB affects performance, yielding a similar sensitivity region, but a much smaller exclusion region (including only $z_0 \gtrsim 10^{-4}$ and $z_{\rm crit} \lesssim -10^{-4}$). The exclusion and sensitivity regions for the $25^{\circ}$ collisions are, again, nearly identical to those for the $10^{\circ}$ collisions.

Based on these results, the first two steps of our pipeline can detect bubble collisions with central modulations of $|z_0| \agt 3 \times 10^{-5}$ {\em or} causal edges of $|z_{\rm crit}| \agt  3 \times 10^{-5}$ with $\theta_{\rm crit} \agt 5^{\circ}$.

{\bf{\em Bayesian analysis:}} Assuming that the bubble collision model likelihood is peaked in the $\nblob$ blobs identified by the needlets, the (unnormalized) posterior probability for the number of detectable collisions, $\nsavge$, given a CMB data set ${\bf d}$ covering a sky fraction $\fsky$, can be approximated as~\cite{Feeney:2010dd}
\begin{eqnarray}
\label{equation:posteriorfinal}
\prob(\nsavge | \data, \fsky) & \propto & 
  \prob(\nsavge) \,  e^{-\fsky \nsavge}
  \sum_{\ns = 0}^\nblob
    \frac{(\fsky \nsavge)^\ns}{\ns!} \\
    & & 
  \sum_{\blob_1, \blob_2, \ldots, \blob_{\ns} = 1}^{\nblob}
    \left[
    \prod_{\src = 1}^{\ns} 
      \rho_{b_s}
    \prod_{i,j = 1}^{\ns} (1 - \delta_{\src_i, \src_j})
    \right],\nonumber
\end{eqnarray}
where $\prob(\nsavge)$ is the prior probability of $\nsavge$ (assumed to be uniform), the pre-factors reflect the fact that the number of collisions present on the observable sky, $\ns$, is the realization of a Poisson-like process (of mean $\fsky \nsavge$), and $\rho_\blob = \prob(\data_\blob | 1)/\prob(\data_\blob | 0)$ is the ratio of the model-averaged likelihoods (or {\em evidences}) of the bubble model and $\Lambda$CDM evaluated within a candidate collision region (with data sub-set $\data_\blob$). The posterior can therefore be built from local measures of how well the data are described by the standard model with and without a single collision template. 

We use the nested sampler MultiNest~\cite{Feroz:2008xx} to calculate the single-blob evidence for the bubble collision model,
\begin{equation}
\label{eq:bubbleev}
\prob(\data_\blob | 1) = \int \diff\model \, \prob(\model) \, \prob(\data_\blob | \model, 1),
\end{equation}
by marginalizing the bubble collision likelihood $\prob(\data_\blob | \model, 1)$ over the model's $n$ parameters $\model$. The (exact) pixel-space likelihood is $\prob(\data_\blob | \model, 1) \propto \exp[-(\data_\blob - \template(\model)) \matr{C}_\blob^{-1} (\data_\blob - \template(\model))^{\rm T}/ 2]$, where $\template(\model)$ is the temperature modulation caused by the collision, and $\matr{C}_\blob$ is the pixel-pixel covariance matrix of the collision region (computationally limited to patches of radius $\lesssim 11^{\circ}$), including cosmic variance given by the best-fit $C_\ell$ as well as the W-band noise and beam from the WMAP 7-year data release. The evidence for $\Lambda$CDM, $\prob(\data_\blob | 0)$, can be calculated by a single evaluation of the bubble likelihood with ${\template(\model)}= {\bf 0}$.

The parameter prior $\prob(\model)$ in Eq.~\ref{eq:bubbleev} is derived from theory, previous experiments and, as only detectable bubble collisions are considered, the limitations of the data-set and pipeline. Lacking a detailed theoretical prediction for the amplitude parameters $\{ z_0, z_{\rm crit} \}$, we assume a uniform prior over the ranges $-10^{-4} \leq \{ z_0, z_{\rm crit} \} \leq 10^{-4}$. Bubble collisions are expected to be distributed isotropically on the CMB sky, so we assume uniform priors on the full ranges of \{$\cos \theta_0, \phi_0$\}. The prior range for {\em detectable} bubble collisions is restricted by CMB power at small scales and computational requirements at large scales. Based on this, we assume uniform priors on $\theta_{\rm crit}$ values between $2^\circ \le \theta_{\rm crit} \le 11^\circ$. 

The evidence ratios found in the end-to-end simulation are bounded by $\ln \rho_\blob \le -6.6$. Evaluating the posterior, it is maximized at $\nsavge = 0$, and we conclude that the residual foregrounds and systematics from the end-to-end simulation do {\em not} provide false support for the collision hypothesis. Evidence ratios for simulated collisions in the needlet and CHT exclusion regions are very large ($\ln \rho_\blob \sim 100$), yielding a full posterior (even for one blob) that clearly favors the  bubble collision hypothesis for a variety of $\nsavge$. For simulated collisions in the sensitivity region, $\rho_\blob$ can cover a wide range ($-7 \lesssim \ln \rho_\blob \lesssim 9$), and does not always yield full-sky posteriors that favor the bubble collision hypothesis. Thus, while we might rule out these features as being due to systematics or foregrounds, better data would be needed to definitively establish the bubble collision hypothesis.

{\bf Analysis of the WMAP 7-year data:} Applying the needlet step of our pipeline to the WMAP 7-year W-band temperature map with the KQ75 mask, we find 15 significant features, four of which intersect the main Galactic mask and are hence discarded (discarding patches from the evidence integrand leads to a conservative lower bound on the integral). The number of features and their significance range is fairly consistent with the end-to-end simulation (we find needlet significances in the range $3.37 \leq S_{jk} \leq 4.58$). The edge-detection results are also consistent with those from the end-to-end simulation, with no clear peaks in the CHT score for any of the significant features. We can therefore rule out the presence of any large-angular-scale bubble collisions with values of $\{z_0, z_{\rm crit}\}$ in the CHT exclusion region of Fig.~\ref{fig-exclusionplots}. The evidence ratios for the 11 significant features are bounded by $\ln \rho \le -3.8$, yielding a posterior for $\nsavge$ that is maximal at $\nsavge = 0$. Therefore, the WMAP 7-year data-set does not favor the bubble collision hypothesis for any value of $\nsavge$; from the posterior we find $\nsavge < 1.6$ at $68 \%$ probability.

Interestingly, however, the Bayesian evidence ratios obtained for four of the features in the WMAP data are systematically larger than expected from false detections, as calibrated using a collision-free end-to-end simulation of the WMAP experiment. Data from the {\em Planck} satellite~\cite{Tauber2010}, which has a resolution three times better than that of WMAP, with an order of magnitude greater sensitivity, will greatly improve the pipeline's diagnostic power. As we have shown, the non-detection of a bubble collision can be used to place constraints on theories giving rise to eternal inflation; however, if a bubble collision is verified by future data, then we will gain an insight not only into our own universe but a multiverse beyond. 

\acknowledgements
We are very grateful to Eiichiro Komatsu and the WMAP Science Team for supplying the end-to-end WMAP simulations used in our null tests. SMF is supported by the Perren Fund. HVP is supported by Marie Curie grant MIRG-CT-2007-203314 from the European Commission, and by STFC and the Leverhulme Trust. MCJ acknowledges support from the Moore foundation. MCJ and HVP thank the Aspen Center for Physics, where this project was initiated, for hospitality.  We acknowledge use of the HEALPix and MultiNest packages and the Legacy Archive for  Microwave Background Data Analysis (LAMBDA).  Support for LAMBDA is provided by the NASA Office of Space Science.

\bibliography{cmbbubbles}

\end{document}